\documentclass[twoside,twocolumn,9pt]{article}
\usepackage{extsizes}
\usepackage[super,sort&compress,comma]{natbib} 
\usepackage[version=3]{mhchem}
\usepackage[left=1.5cm, right=1.5cm, top=1.785cm, bottom=2.0cm]{geometry}
\usepackage{balance}
\usepackage{times,mathptmx}
\usepackage{sectsty}
\usepackage{graphicx} 
\graphicspath{{./}{./figs/}}
\usepackage{lastpage}
\usepackage[format=plain,justification=justified,singlelinecheck=false,font={stretch=1.125,small,sf},labelfont=bf,labelsep=space]{caption}
\usepackage{float}
\usepackage{fancyhdr}
\usepackage{fnpos}
\usepackage[english]{babel}
\usepackage{array}
\usepackage{droidsans}
\usepackage{charter}
\usepackage[T1]{fontenc}
\usepackage[usenames,dvipsnames]{xcolor}
\usepackage{setspace}
\usepackage[compact]{titlesec}
\usepackage{pdfpages}

\usepackage{epstopdf}

\definecolor{cream}{RGB}{222,217,201}

\usepackage{cleveref}
\Crefformat{figure}{Fig.~#2#1#3}
\usepackage{amssymb}

\usepackage{xspace}
\newcommand{\lenunit}{\ensuremath{\ \mu \textrm{m}{\xspace}}}
\newcommand{\densunit}{\ensuremath{\ \mu \textrm{m}^{-2}{\xspace}}}
\newcommand{\rateunit}{\ensuremath{{\xspace}\textrm{s}^{-1}{\xspace}}}

\newcommand{\msd}{\ensuremath{\textrm{MSD}(t_f,\Delta)}}

\begin{document}

\pagestyle{fancy}
\thispagestyle{plain}


\makeFNbottom
\makeatletter
\renewcommand\LARGE{\@setfontsize\LARGE{15pt}{17}}
\renewcommand\Large{\@setfontsize\Large{12pt}{14}}
\renewcommand\large{\@setfontsize\large{10pt}{12}}
\renewcommand\footnotesize{\@setfontsize\footnotesize{7pt}{10}}
\makeatother

\renewcommand{\thefootnote}{\fnsymbol{footnote}}
\renewcommand\footnoterule{\vspace*{1pt}%
\color{cream}\hrule width 3.5in height 0.4pt \color{black}\vspace*{5pt}} 
\setcounter{secnumdepth}{5}

\makeatletter 
\renewcommand\@biblabel[1]{#1}            
\renewcommand\@makefntext[1]%
{\noindent\makebox[0pt][r]{\@thefnmark\,}#1}
\makeatother 
\renewcommand{\figurename}{\small{Fig.}~}
\sectionfont{\sffamily\Large}
\subsectionfont{\normalsize}
\subsubsectionfont{\bf}
\setstretch{1.125} 
\setlength{\skip\footins}{0.8cm}
\setlength{\footnotesep}{0.25cm}
\setlength{\jot}{10pt}
\titlespacing*{\section}{0pt}{4pt}{4pt}
\titlespacing*{\subsection}{0pt}{15pt}{1pt}


\makeatletter 
\newlength{\figrulesep} 
\setlength{\figrulesep}{0.5\textfloatsep} 

\newcommand{\topfigrule}{\vspace*{-1pt}%
\noindent{\color{cream}\rule[-\figrulesep]{\columnwidth}{1.5pt}} }

\newcommand{\botfigrule}{\vspace*{-2pt}%
\noindent{\color{cream}\rule[\figrulesep]{\columnwidth}{1.5pt}} }

\newcommand{\dblfigrule}{\vspace*{-1pt}%
\noindent{\color{cream}\rule[-\figrulesep]{\textwidth}{1.5pt}} }

\makeatother

\twocolumn[
  \begin{@twocolumnfalse}
\sffamily

\noindent\LARGE{\textbf{Design principles for selective self-assembly of active networks}} \\
\vspace{0.3cm} \\

\noindent\large{Simon L. Freedman,\textit{$^{a}$} Glen M. Hocky,\textit{$^{b}$} Shiladitya Banerjee,\textit{$^{c}$} and Aaron R. Dinner\textit{$^{b}$}} \\

\noindent\normalsize{Living cells dynamically modulate the local morphologies of their actin cytoskeletons to perform biological functions, including force transduction, intracellular transport, and cell division. 
A major challenge is to understand how diverse structures of the actin cytoskeleton are assembled from a limited set of molecular building blocks. 
Here we study the spontaneous self-assembly of a minimal model of cytoskeletal materials, consisting of semiflexible actin filaments, crosslinkers, and molecular motors. 
Using coarse-grained simulations, we demonstrate that by changing concentrations and kinetics of crosslinkers and motors we can generate three distinct structural phases of actomyosin assemblies: bundled, polarity-sorted, and contracted. 
We introduce new metrics to distinguish these structural phases and demonstrate their functional roles. 
We find that the binding kinetics of motors and crosslinkers can be tuned to optimize contractile force generation, motor transport, and mechanical response. 
By quantitatively characterizing the relationships between modes of cytoskeletal self-assembly, the resulting structures, and their functional consequences, our work suggests new principles for the design of active materials.  
}\\

 \end{@twocolumnfalse} \vspace{0.6cm}

  ]

\renewcommand*\rmdefault{bch}\normalfont\upshape
\rmfamily
\section*{}
\vspace{-1cm}


\footnotetext{\textit{$^{a}$~Department of Physics, The University of Chicago, 929 East 57th Street, Chicago, IL 60637, USA}}
\footnotetext{\textit{$^{b}$~James Franck Institute \& Department of Chemistry, The University of Chicago, 929 East 57th Street, Chicago, IL 60637, USA, Chicago, IL, USA; Email: dinner@uchicago.edu}}
\footnotetext{\textit{$^{c}$~Department of Physics and Astronomy, University College London, Gower Street, London, WC1E-6BT E-mail: shiladitya.banerjee@ucl.ac.uk}}




\section{Introduction} 
Mechanical functions of living cells are determined by dynamic restructuring of the actin cytoskeleton, a highly conserved cellular machinery composed of filamentous actin (F-actin), myosin molecular motors, and crosslinking proteins~\cite{murrell2015}. 
An enormous variety of F-actin binding proteins with diverse physico-chemical properties~\cite{michelot2011} can combine with F-actin to assemble function-specific cellular structures.
Spatiotemporal control over these structures is essential for coordinated force generation during cell migration~\cite{pollard2003,lomakin2015}, cell adhesion~\cite{parsons2010}, cytokinesis~\cite{sedzinski2011}, and intracellular transport~\cite{munro2004,tabei2013}. 
A quantitative understanding of how diverse cytoskeletal structures are assembled from a limited set of molecular building blocks presents an outstanding challenge at the interface of soft matter physics and cell biology.

Given the many interconnected processes within cells, it is experimentally difficult to controllably study how variations in molecular-scale properties affect emergent actin network structures and function.  
This issue can be overcome by studying {\em in vitro} reconstitutions, which have revealed how biochemical compositions and relative amounts of actin filaments and molecular motors determine network architectures and mechanical properties~\cite{bendix2008,kohler2011,alvarado2013, murrell2012, murrell2014, ennomani2016,linsmeier2016,chugh2017,stam2017}.
Such studies have shown that the types and concentrations of crosslinkers can be used to tune the length scale \cite{murrell2012,stam2017} and shape \cite{stam2017} of microscopic contractile deformations, as well as bulk mechanical responses of actin networks \cite{gardel2004,kasza2009,freedman2017,schmoller2009,weirich2017,lieleg2009} and motor transport on them \cite{scholz2018}.


Simulations complement experiments by allowing both precise control of the physical properties of constituents and examination of microscopic mechanisms.  
The foci of existing simulations have mirrored the experiments described above.
\begin{itemize}
\item
Simulations have been used to determine the  conditions necessary for actomyosin networks to be contractile due to the formation of force chains~\cite{dasanyake2011}, contractile due to sliding and buckling of semi-flexible filaments~\cite{ennomani2016,stam2017}, or extensile due to rigid filament sliding~\cite{belmonte2017,stam2017}.
\item Simulations have been used to determine shear moduli for networks of semi-flexible filaments and crosslinkers of varying stiffnesses~\cite{head2003,kim2009,freedman2017}
\item Simulations have been used to predict network characteristics that lead to anomalous motor transport dynamics~\cite{head2011,scholz2016}.
\end{itemize}

The simulation studies above provide insight into specific aspects of cytoskeletal networks (contractility, mechanical response, and transport) for assembled networks.  
They do not consider dramatic structural rearrangements in response to internal and external forces, and it is difficult to compare the functional consequences of specific structures between studies.
Indeed, while the structural phase diagram of filaments with passive crosslinkers has been mapped and shown to exhibit homogeneous isotropic gel, bundled, clustered, and lamellar phases \cite{cyron2013}, to the best of our knowledge, a corresponding study of active materials has not been performed previously. 
It remains to determine the structures accessible to mixtures of filaments, crosslinkers, and motors, the conditions under which they are formed, and the interplay of network structure and function.

In this paper, we map the non-equilibrium structural phases of networks consisting of F-actin, crosslinkers, and motors. 
We observe homogeneous, bundled, contracted, and polarity-sorted  networks within a single unified model, and characterize the parameters that control their assembly.  
In doing so, we introduce order parameters that can classify the extent to which networks contract into dense aggregates, sort F-actin by polarity, or bundle filaments for force propagation. 
Using these metrics, we demonstrate how networks can be tuned for specific mechanical functionalities, by systematic variations in network composition. 
New insights offered by our work include that varying kinetic properties of actin binding proteins and filament length effect network structure non-monotonically, implying optimal values for these parameters, and how different network structures set time scales of motor transport.

\section{Results \& Discussion}
\subsection{Coarse-grained model\label{secn:model}}
\begin{figure}[t]
  \centering
  \includegraphics[width=\columnwidth]{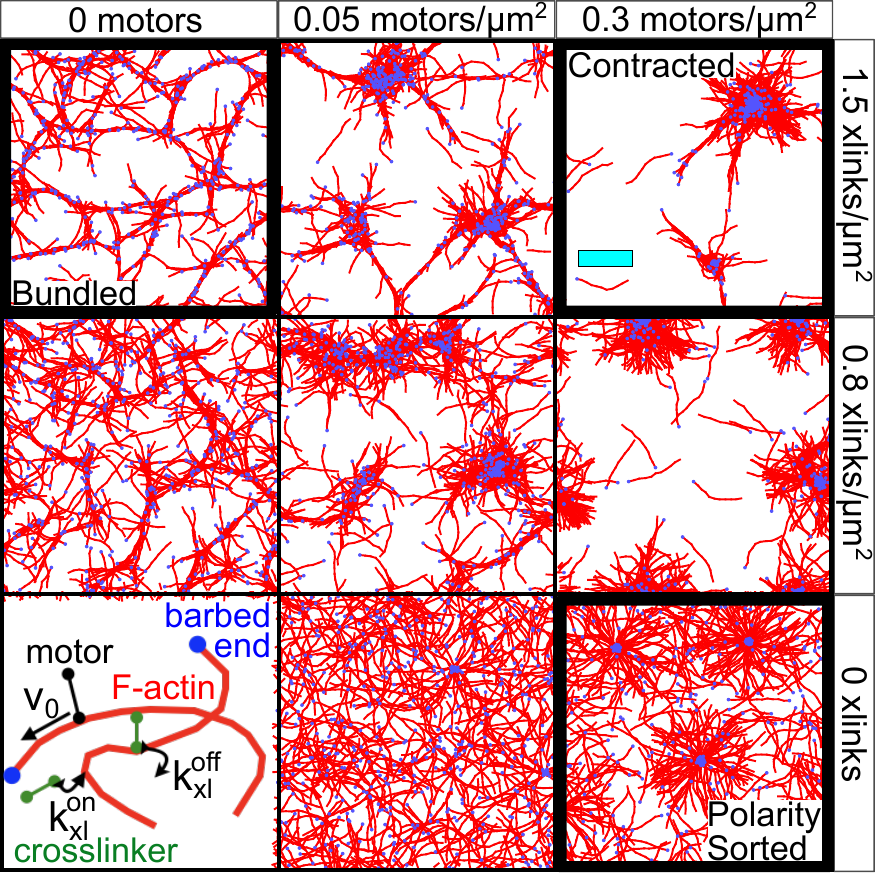}
  \caption{
  \label{fig:structs}
  Network structures.  
  (Lower left) Schematic of the model \cite{freedman2017} with F-actin (red), crosslinkers (green), and motors (black). 
Bound motors walk towards F-actin barbed ends (blue).
  (Remaining panels) Network structures at $400$~s for indicated motor and crosslinker densities.
These include a bundled network (upper left) formed by filaments and crosslinkers, a polarity-sorted network (lower right) formed by filaments and motors, and a contracted network (upper right) formed from filaments, crosslinkers, and motors. 
For clarity, only the actin filaments are shown; the motors and crosslinkers are shown in Fig. S1.
Cyan scale bar represents $10\lenunit$.
}
\end{figure}

To study the spontaneous self-assembly of cytoskeletal structures at experimentally relevant length and time scales (microns and minutes), we use AFINES, a simulation framework we recently developed \cite{freedman2017}. 
In brief, actin filaments are modeled as polar worm-like chains (represented by beads connected by springs) with defined barbed and pointed ends (\Cref{fig:structs}); crosslinkers are modeled as linear springs with ends (heads) that can stochastically bind and unbind from F-actin via a kinetic Monte Carlo procedure that preserves detailed balance; molecular motors are modeled as active crosslinkers such that once bound, they walk toward the barbed ends of filaments at a load-dependent speed.  
We use Brownian dynamics to evolve the positions of constituents in 2D.  
While other similar simulation frameworks exist~\cite{nedelec2007,popov2016}, the unique implementation of detailed balance preservation in AFINES allows one to differentiate between passive systems without motors that assemble thermally, and active systems that break detailed balance, e.g. using motors which hydrolyze ATP to promote unidirectional motion.
Restriction to 2D is consistent with the fact that {\it in vitro} reconstitutions of actomyosin networks are nearly flat \cite{schmoller2009,murrell2012,stam2017}.  
To enable rearrangement in 2D, we neglect excluded volume, which is reasonable since the F-actin density in our simulations is well below the isotropic to nematic transition, so network connectivity dominates the dynamics.
The model is described in detail in Section S1, and Table S1 lists all simulation parameters.

\subsection{Self-assembly and characterization of actin network structures} 
We observe three distinct network architectures formed from initially disordered mixtures of F-actin, motors, and crosslinkers:  bundled, polarity-sorted, and contracted. 
Examples are shown for simulations of 500 10 $\mu$m long filaments in \Cref{fig:structs}. 
When F-actin is mixed with crosslinkers, thick bundles form and intersect to yield a well-connected mesh. 
When F-actin is mixed with motors, barbed ends aggregate to form a polarity-sorted network.
Combining F-actin with both motors and crosslinkers results in macroscopic contraction of the filaments into dense and disconnected aggregates.

\begin{figure}[t]
  \centering
  \includegraphics[width=\columnwidth]{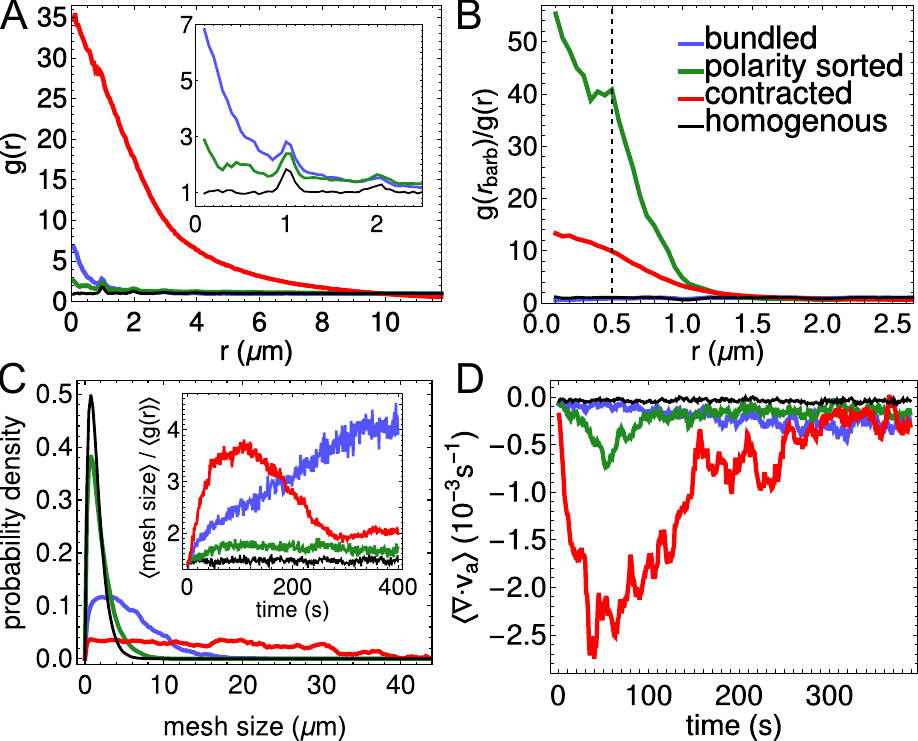}
  \caption{
    \label{fig:quants}
Order parameters.  
Bundled, polarity-sorted, and contracted cases correspond to the conditions highlighted in \Cref{fig:structs}; the homogeneous case has no motors or crosslinkers. 
(A) Radial distribution function of actin, $g(r)$.
(B) Radial distribution function of F-actin barbed ends, normalized by $g(r)$; the dashed line marks the motor rest length, $0.5\lenunit$.
(C) Mesh size distribution.
Inset: Evolution of the average mesh size, normalized by $\langle g(r)\rangle$.
(D) Spatially averaged divergence (Section S2).
 All averages are over the final 50 s of 5 simulations of 400 s. 
}
\end{figure}

To systematically explore how varying the properties of the network constituents affects structure formation, we introduce physical order parameters that characterize each of the observed structural phases.
We compute the spatial extent of F-actin aggregation using the radial distribution function, 
\begin{equation}
 g(r) = P(r)/(2\pi r \delta r\rho_a)
  \label{eqn:gr}
\end{equation}
where $P(r)$ is the probability that two actin beads are separated by a distance in the range $[r, r+\delta r]$ (here, $\delta r =0.05$ $\mu\mathrm{m}$), and $\rho_a$ is the number density of actin beads. 
For a homogeneous network, $g(r)\approx 1$ at all distances (\Cref{fig:quants}A; the small peaks at integer $r$ arise from the spacing of beads within actin filaments).
In contrast, for contracted networks,  $g(r)\gg 1$ for $r<10\lenunit$, indicating F-actin exceeds the bulk density. 

While \Cref{fig:quants}A shows that actin filaments are nearly uniformly distributed in a polarity-sorted network, \Cref{fig:structs} indicates that their barbed ends are concentrated. 
To quantify their aggregation specifically, we compute the ratio $g(r_{barb})/g(r)$, where $r_{barb}$ is the distance between barbed ends. 
\Cref{fig:quants}B shows that in polarity-sorted networks F-actin barbed ends aggregate (and have a secondary peak at $0.5\lenunit$, the rest length of motors). 
In contracted networks, barbed ends also aggregate to a higher degree than in bundled networks, indicating a degree of polarity sorting. 

\begin{figure}[h]
  \centering
  \includegraphics[width=0.8\columnwidth]{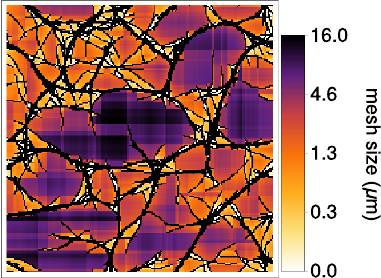}
  \caption{
    \label{fig:mesh}
    Example of mesh size calculation described in the text. 
    Actin is shown in black (compare with bundled structure in Fig.\ \ref{fig:structs}) and the size of the local mesh corresponds to the depth of color, as indicated by the scale.
  }
\end{figure}

\Cref{fig:quants}A indicates that bundled networks aggregate at smaller length scales, $\sim0.1\lenunit$, corresponding to the crosslinker rest length $l_{xl}=0.15\lenunit$.
To quantify the degree of bundling, and to distinguish it from contractility, we measure the distribution of network pore sizes by a procedure that is similar in spirit but simpler than that in Ref.~\citenum{mickel2008}. 
Namely, we grid the simulation box into $(0.25\lenunit)^2$ bins and compute how many filaments pass through each.
For each empty bin, we determine the lengths of the contiguous vertical and horizontal stretches of empty bins that intersect it (\Cref{fig:mesh}). 
We average these lengths over all empty bins to obtain an average mesh size for each structure.
This procedure can be used for analysis of experimental images, in addition to the simulation structures in the present study.

In \Cref{fig:quants}C, the distributions of mesh sizes for polarity-sorted and homogeneous networks are similar, indicating that the former does not coarsen significantly.  
The bundled and contracted networks exhibit larger pore sizes; indeed, contracted networks exhibit pore sizes spanning the simulation region, indicating that the network has ripped apart.  
We can distinguish these cases by normalizing the mesh size by $\langle g(r)\rangle=(1/R)\int_0^{R}{g(r)dr}$, (here, $R=10\lenunit$ is the approximate size of a contracted aggregate under our maximally contractile conditions, as shown in \Cref{fig:quants}A; see Fig. S2 for further details) which quantifies the extent of aggregation. 
The inset shows that, while the contracted networks initially bundle, at long times this effect is small compared to aggregation.
In contrast, bundled networks have a continuously increasing normalized mesh size (\Cref{fig:quants}C, inset); we thus have a metric for the degree of bundling.

To examine the relationship between actin network structure and contractility, we use the divergence of actin's velocity field,  $\langle \nabla\cdot {\bf v} \rangle$, where $\langle ... \rangle$ indicates spatial averaging (Section S2) \cite{freedman2017}. 
As shown in \Cref{fig:quants}D, $\langle \nabla\cdot {\bf v} \rangle$ becomes significantly more negative for aggregating networks than for bundling or polarity-sorted networks.  
Comparison with \Cref{fig:quants}A shows that extensive contractility is associated with large $g(r)$.

\begin{figure}[ht]
  \centering
  \includegraphics[width=\columnwidth]{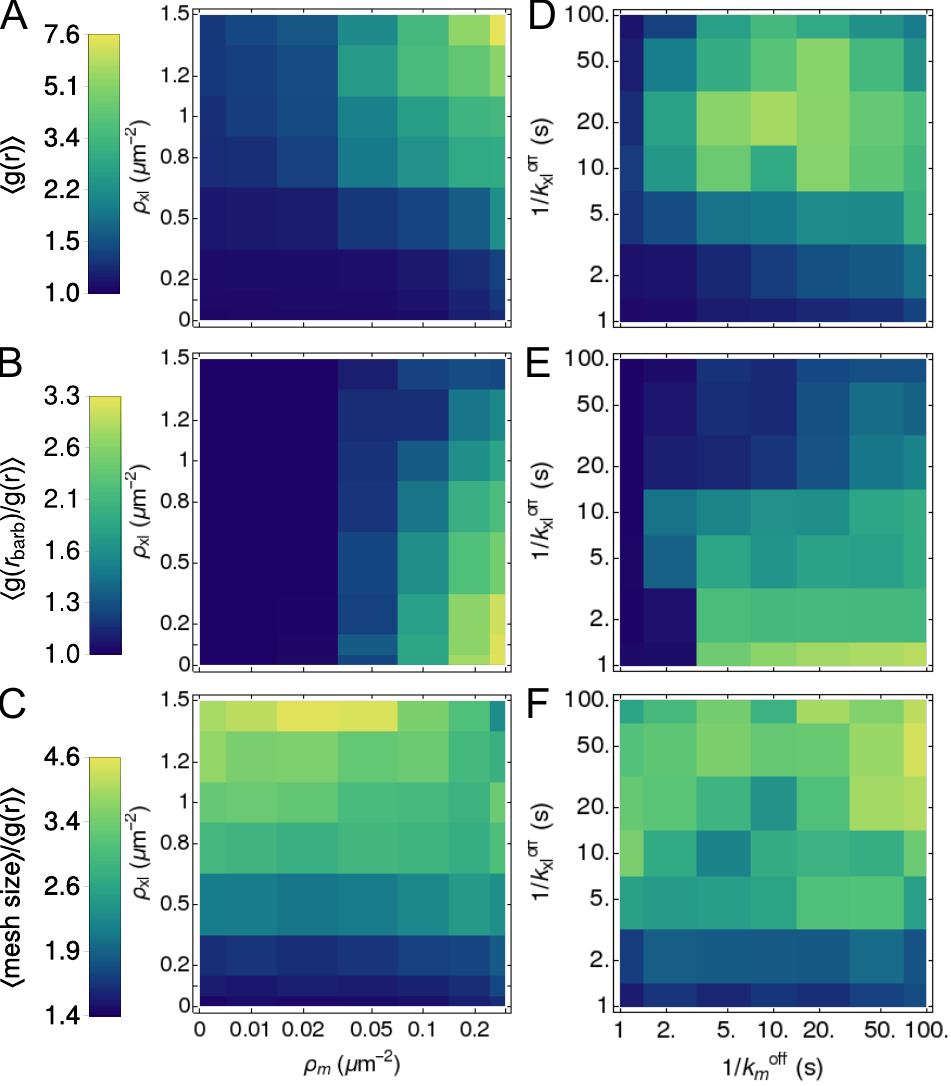} 
  \caption{
    \label{fig:phases}
    Maps of network properties.  
    Colors indicate the values of order parameters characterizing network contractility (A,D), polarity sorting (B,E), and bundling (C,F), at constant filament length, $L=10\lenunit$, binding affinity $k_m^{\rm{off}}=k_{xl}^{\rm{off}}=0.1\ \rateunit$ (left), and densities $\rho_m=0.2\densunit$, $\rho_{xl}=1\densunit$ (right; structures shown in Fig. S4).
    Averages are over the last 50 s of 5 simulations of 400 s; order parameters that are functions of distance, e.g., $g(r)$, are integrated over $0<r\leq 10\lenunit$.
  }
\end{figure}

Using these order parameters, we map the structural phase space of actomyosin networks as functions of motor and crosslinker densities and their binding affinities (\Cref{fig:phases}).  
Consistent with \Cref{fig:structs}, networks are contracted when motor and crosslinker densities are high (\Cref{fig:phases}A), polarity-sorted when only motor density is high (\Cref{fig:phases}B), and bundled when crosslinker density is high (\Cref{fig:phases}C).  
Interestingly, while high motor densities inhibit bundling, a small population of motors ($\rho_m\approx0.02\densunit$) enhances filament bundling.

\subsection{Non-monotonic trends in binding kinetics and filament length\label{secn:kofflen}} 
\begin{figure}[ht]
  \centering
  \includegraphics[width=0.95\linewidth]{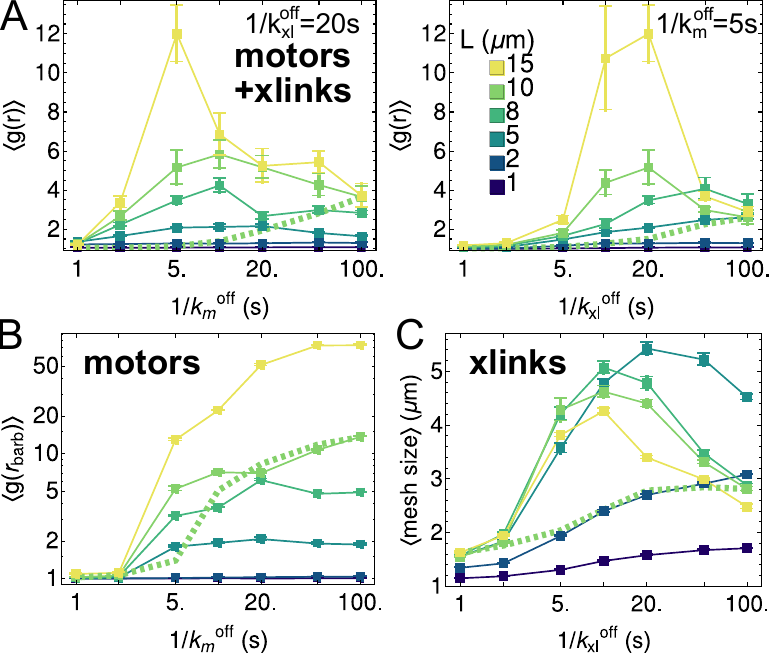} 
  \caption{
    \label{fig:koff}
    Effects of varying motor and crosslinker affinities for filaments of different lengths, $L$, at $t_f=400$ s (filled squares).
    (A) At $\rho_m=0.2\densunit$ and $\rho_{xl}=1\densunit$, contractility is maximized at finite values of  $k^{\rm{off}}_{m}$ and $k^{\rm{off}}_{xl}$ for $L\ge5\lenunit$.
    (B) Without crosslinkers, polarity sorting generally increases with higher $L$ and higher $1/k^{\rm{off}}_{m}$.
    (C) Without motors, there is an optimal $k^{\rm{off}}_{xl}$ and $L$ for maximizing the mesh size. Normalizing quantities in (B) and (C)  with respect to $\langle g(r)\rangle$ does not change these conclusions (Fig. S3).  Error bars are standard error of the mean.
    The green dashed lines show how order parameters vary for one length, $L=10\lenunit$, if the corresponding green squares are rescaled such that $t_f=4/k_{xl(m)}^{\rm{off}}$.
  }
\end{figure}

Finally, we modulate molecular-level interaction parameters between F-actin and its binding partners (crosslinkers, motors) to dissect their relative roles in building different structures. 
These parameters are hard to change independently in experiment. 
At fixed motor and crosslinker densities, we find that cytoskeletal structures can be tuned by varying the dissociation constants, $k_{m(xl)}^{\rm{off}}$ (\Cref{fig:phases}D-F).
The trends are non-monotonic, in contrast to those in \Cref{fig:phases}A-C.
In particular, contraction is highest for intermediate values of $k_{m,xl}^{\rm{off}}$ (\Cref{fig:phases}D), 
and bundling is highest for low values of $k_{xl}^{\rm{off}}$ with low or high $k_m^{\rm{off}}$ (\Cref{fig:phases}F).
Notably, this non-monotonic trend only arises for a fixed simulation time, $t_F$. 
For a fixed value of $t_Fk^{\rm{off}}_{m(xl)}$, changing $1/k^{\rm{off}}_{m(xl)}$ modulates structure formation in a monotonic manner, akin to changing $\rho_{m(xl)}$ (dashed lines in \Cref{fig:koff}).  
The non-monotonic trends are important, however, for understanding how structures form in the presence of competing kinetic processes, such as actin filament turnover \cite{munro2004,fritzsche2016}.

As the length ($L$) of F-actin varies considerably within cells \cite{mohapatra2016}, we tested how modulating $L$ and $k^{\rm{off}}_{m(xl)}$ in tandem affects structure formation.
In \Cref{fig:koff}A, we show that increasing $L$ favors aggregation by increasing network connectivity; short filaments ($L<5\lenunit$) do not organize.
As in \Cref{fig:phases}E, the dependence of aggregation on binding affinity is non-monotonic whenever there is significant aggregation (i.e., $L\ge5\lenunit$).
In \Cref{fig:koff}B, we show that for networks with only motors, at low $k_m^{\rm{off}}$, increasing filament length promotes polarity sorting. 
By contrast, at high $k_m^{\rm{off}}$,  short lifetimes of motor attachment suppress polarity sorting.
As evident from the representative network structures in Fig. S5, shortening the filaments also suppresses polarity sorting.  
Both the large number of filaments and their rapid diffusion favor mixing over sorting. 

For networks with only crosslinkers, the mesh size (\Cref{fig:koff}C) is non-monotonic with respect to both filament length and crosslinker affinity.
Low crosslinker affinity and short filament lengths prevent forming stably crosslinked networks (Fig. S6), again due to mixing.
Conversely, assemblies with high crosslinker affinity or long filament lengths form crosslinked networks, but they rearrange slowly, so further coarsening is impeded, and the mesh size remains small.
As these non-monotonic trends only occur for $L>5$ $\mu$m, they are more likely to impact structures with longer actin filaments found in budding yeast~\cite{chesarone2011}, stereocilia~\cite{lin2005}, filopodia, or {\em in vitro} reconstituted networks~\cite{weirich2017,stam2017}. Structures with shorter filaments, as found in lamellipodia or the actin cortex ($<2$ $\mu$m)~\cite{salbreux2012,chugh2017}, are less likely to have a finite binding affinity that maximizes contractility.

\subsection{Network structure tunes transport and force propagation}
\begin{figure}[t]
  \centering
  \includegraphics[scale=1.2]{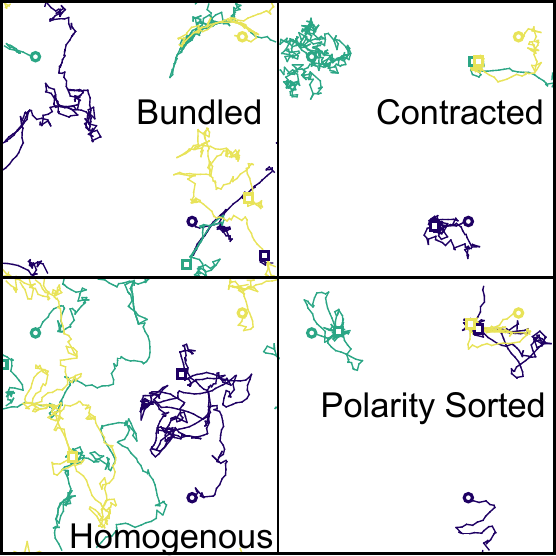}
  \caption{Trajectories of three motors (each a different color) on fixed actin networks.
    Actin filaments for bundled, polarity sorted and contracted networks are shown in \Cref{fig:structs}; homogenous network is shown in Fig. S1.
    Open circles show initial position of each motor, and open squares show their final position after $500$ timesteps.}
  \label{fig:traj}
\end{figure}

\begin{figure}[h]
  \centering
  \includegraphics{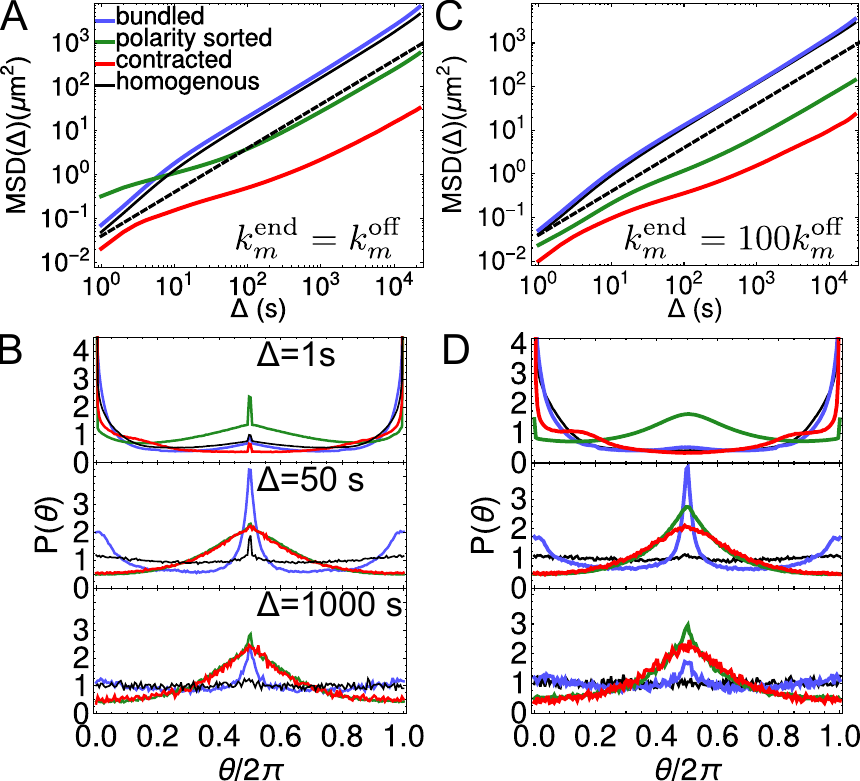}
  \caption{Transport properties of motors on fixed actin structures.
    All averages are over $n_m=1,250$ motors, $t_f=25,000$ s, and $n_s=5$ congruent strutures.
    (A) Mean squared displacement for motors on structures.
    Black dashed line shows diffusive behavior, $\msd\propto\Delta$.
    (B) Distributions of angles between subsequent motor displacements for different lag times, $\Delta$\cite{burov2013}.
    (C-D) Same as (A-B) but for motors with increased off-rate at filament barbed ends to surpress the spikes at $\theta=\pi$ in (B).
    All lag times for (B) and (D) are shown in Fig. S7.
  }
  \label{fig:msd}
\end{figure}

While the structures of contracted, polarity-sorted, and bundled networks clearly differ, their consequences for biophysical functions are not immediately apparent. 
To determine how these structures influence motor transport \cite{brawley2009, kohler2011}, we fix the actin structures and follow the dynamics of the motors (\Cref{fig:traj}).  
This facilitates obtaining well-converged statistics for the dynamics and interpretation in terms of individual trajectories.
In \Cref{fig:msd}A, we plot their mean-squared displacement,
\begin{equation}
  \msd=\frac{1}{t_f-\Delta}\int_0^{t_f-\Delta}{[\vec{v}_m(t, \Delta)]^2 dt}
  \label{eqn:msd}
\end{equation}
where $t_f$ is the length of the trajectory and $\vec{v}_m(t, \Delta)=\vec{r}_m(t+\Delta)-\vec{r}_m(t)$ is the displacement of a motor with center of mass position $\vec{r}_m(t)$ at time $t$ after a lag of $\Delta$. 
While the scaling of the mean-squared displacement is consistent with simple diffusion (\Cref{fig:msd}A), sample trajectories (\Cref{fig:traj}) indicate that motors in contracted and polarity-sorted structures spend significant amounts of time trapped in aggregates of barbed ends. 

We quantify caging using a previously defined metric that can distinguish different kinds of motion~\cite{burov2013}:  we compute the distribution of angles, $\theta$, between consecutive displacement vectors $\vec{v}_m(t,\Delta)$ and $\vec{v}_m(t+\Delta, \Delta)$, at different values of $\Delta$.
We performed this measurement with the motors described above (\Cref{secn:model}), as well as ones that detach faster from the barbed ends of filaments; i.e., at the barbed end they had a detachment rate of $k_m^{\rm{end}}=100k_m^{\rm{off}}$~\cite{freedman2017}.
This second set of measurements was done to ascertain that the observed motor dynamics were a consequence of network structure, and not a consequence of dwelling at the barbed ends of the filaments, given that we assume a uniform detachment rate. 

We find that for all structures, there is at least one time scale in which the distribution has a broad peak at $\theta=\pi$  (\Cref{fig:msd}B), indicating that motors are reversing direction, consistent with confinement \cite{burov2013}. 
Motors with $k_m^{\rm{end}}=k_m^{\rm{off}}$ exhibit an additional spike in their distribution at $\theta=\pi$ due to barbed ends acting as local attractors, between which motors move back and forth.
However, this spike is suppressed by increasing motor end detachment rate, $k_m^{\rm{end}}=100k_m^{\rm{off}}$ and does not impact the diffusive or caging behavior (\Cref{fig:msd}C-D).
For polarity-sorted networks barbed ends are most tightly aggregated, and thus motors exhibit caging at all time scales measured.
Contracted networks are partially polarity-sorted, so filaments can direct motors both in and out of aggregates, making the caging more spatially extended.  
Because it takes longer to explore the extended length scale, the caging manifests only at $\Delta\gtrsim 50$ s.
Bundled networks show caging effects most prominently at intermediate time scales.
In this case, the caging corresponds to motors cycling between oppositely oriented filaments, which can give rise to apparently glassy dynamics \cite{scholz2016}.
Consistent with our results, it was recently shown that actin networks with different structures result in different angle distributions \cite{scholz2018}.  
However, the caging manifested in the experiments at shorter times ($\sim 0.1-1$ s), and certain networks supported simultaneous stable peaks at $\theta=0$ and $\theta=\pi$ (see also \cite{burov2013}). 
A key feature of the experiments that is not represented in the present model is that the myosin minifilaments in the experiments have many heads \cite{burov2013,scholz2016,scholz2018}, and this was previously shown to be important in producing those observed glassy dynamics \cite{scholz2016}.

\begin{figure}[H]
  \centering
  \includegraphics[width=\columnwidth]{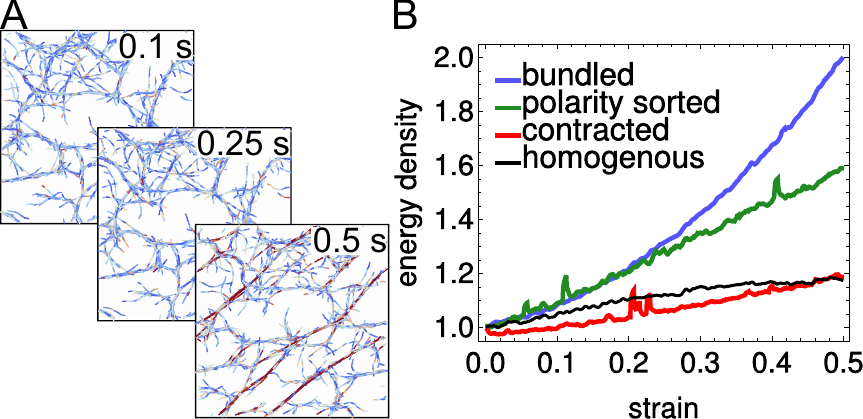}
  \caption{
    \label{fig:shear}
Elasticity of network structures in \Cref{fig:structs}.
(A) Shearing a bundled network for $0.5$ s at a strain rate of $\dot{\gamma}=1 $/s. 
Depth of color indicates stretch on filament. 
(B) Network elasticity as measured by the dependence of the normalized energy density on the strain. 
Fitting unnormalized curves to a quadratic function yields a shear modulus of $G\approx0.66$ Pa for the bundled network (within the range of experimentally measured values for actin-filamin networks~\cite{schmoller2009}) and $G\approx0$ for the other networks. 
}
\end{figure}

Next, we evaluated how structural rearrangements in actin networks affect their ability to propagate mechanical forces over long length scales. 
To this end, we subjected the final network configuration to a shear strain of magnitude $\gamma=0.5$ (\Cref{fig:shear}A, algorithm described in Section S3) and measured the resulting strain energy. 
In \Cref{fig:shear}B, we show the strain dependence of the total strain energy density, $w = (U_f+U_m+U_{xl})/V$, where $U_f$, $U_m$, and $U_{xl}$ are the potential energies of the F-actin, motors, and crosslinkers, respectively, (Section S1) and $V=250\lenunit^3$ is the simulation volume, assuming a thickness of $0.1\lenunit$. 
Bundled networks exhibit a quadratic dependence on strain, indicating a solid-like material response. 
By contrast, contracted and polarity-sorted networks are fragile, with linear dependences of energy density on strain.

\section{Conclusions}
We have shown how modulating the abundance and properties of cytoskeletal constituents can tune emergent network structures.
For finite times, we find that motor and crosslinker binding affinities, as well as filament lengths, have optimal values for maximizing contractility, bundling, and polarity sorting.
While there are still many unanswered questions on the self-organized behavior of cytoskeletal materials, our work takes an important step forward by linking kinetics to the selection of network structures with specific mechanical functions. 
We expect that the order parameters that we introduced can also be applied to interpretation of experimental images.
Further simulations using our modeling framework can shed light on the structures accessible to mixtures with multiple types of crosslinkers, filaments, and motors, as well as structure-function relations in other active polymer assemblies, including networks of microtubules, kinesin, and dynein~\cite{sanchez2012,foster2015}.
From the perspective of materials design, our work demonstrates how a limited set of molecular building blocks can self-assemble diverse active materials. 
This presents the prospect that theory and simulation can be used to guide the design of active biomimetic materials with desired collective mechanical properties.

\section{Conflicts of Interest}
There are no conflicts of interest to declare. 
\section{Acknowledgments}
We thank M. Gardel, J. Weare, S. Stam, and K. Weirich for helpful conversations and anonymous reviewers for insightful comments and suggestions on the manuscript.
This research was supported in part by the University of Chicago Materials Research Science and Engineering Center (NSF Grant No. 1420709).
S.L.F. was supported by the DoD through the NDSEG Program.
G.M.H. was supported by an NIH Ruth L. Kirschstein NRSA award (1F32GM113415-01). S.B. was supported by a UCL Strategic Fellowship.
Simulations resources were provided by the Research Computing Center at the University of Chicago and National Institutes of
Health (NIH) Grant No. 5 R01 GM109455-02.

\balance



\begin{mcitethebibliography}{43}
\providecommand*{\natexlab}[1]{#1}
\providecommand*{\mciteSetBstSublistMode}[1]{}
\providecommand*{\mciteSetBstMaxWidthForm}[2]{}
\providecommand*{\mciteBstWouldAddEndPuncttrue}
  {\def\EndOfBibitem{\unskip.}}
\providecommand*{\mciteBstWouldAddEndPunctfalse}
  {\let\EndOfBibitem\relax}
\providecommand*{\mciteSetBstMidEndSepPunct}[3]{}
\providecommand*{\mciteSetBstSublistLabelBeginEnd}[3]{}
\providecommand*{\EndOfBibitem}{}
\mciteSetBstSublistMode{f}
\mciteSetBstMaxWidthForm{subitem}
{(\emph{\alph{mcitesubitemcount}})}
\mciteSetBstSublistLabelBeginEnd{\mcitemaxwidthsubitemform\space}
{\relax}{\relax}

\bibitem[Murrell \emph{et~al.}(2015)Murrell, Oakes, Lenz, and
  Gardel]{murrell2015}
M.~Murrell, P.~W. Oakes, M.~Lenz and M.~L. Gardel, \emph{Nat. Rev. Mol. Cell
  Biol.}, 2015, \textbf{16}, 486\relax
\mciteBstWouldAddEndPuncttrue
\mciteSetBstMidEndSepPunct{\mcitedefaultmidpunct}
{\mcitedefaultendpunct}{\mcitedefaultseppunct}\relax
\EndOfBibitem
\bibitem[Michelot and Drubin(2011)]{michelot2011}
A.~Michelot and D.~G. Drubin, \emph{Curr. Biol.}, 2011, \textbf{21},
  R560--R569\relax
\mciteBstWouldAddEndPuncttrue
\mciteSetBstMidEndSepPunct{\mcitedefaultmidpunct}
{\mcitedefaultendpunct}{\mcitedefaultseppunct}\relax
\EndOfBibitem
\bibitem[Pollard and Borisy(2003)]{pollard2003}
T.~D. Pollard and G.~G. Borisy, \emph{Cell}, 2003, \textbf{112}, 453--465\relax
\mciteBstWouldAddEndPuncttrue
\mciteSetBstMidEndSepPunct{\mcitedefaultmidpunct}
{\mcitedefaultendpunct}{\mcitedefaultseppunct}\relax
\EndOfBibitem
\bibitem[Lomakin \emph{et~al.}(2015)Lomakin, Lee, Han, Bui, Davidson, Mogilner,
  and Danuser]{lomakin2015}
A.~J. Lomakin, K.-C. Lee, S.~J. Han, D.~A. Bui, M.~Davidson, A.~Mogilner and
  G.~Danuser, \emph{Nat. Cell Biol.}, 2015, \textbf{17}, 1435\relax
\mciteBstWouldAddEndPuncttrue
\mciteSetBstMidEndSepPunct{\mcitedefaultmidpunct}
{\mcitedefaultendpunct}{\mcitedefaultseppunct}\relax
\EndOfBibitem
\bibitem[Parsons \emph{et~al.}(2010)Parsons, Horwitz, and
  Schwartz]{parsons2010}
J.~T. Parsons, A.~R. Horwitz and M.~A. Schwartz, \emph{Nat. Rev. Mol. Cell
  Biol.}, 2010, \textbf{11}, 633--643\relax
\mciteBstWouldAddEndPuncttrue
\mciteSetBstMidEndSepPunct{\mcitedefaultmidpunct}
{\mcitedefaultendpunct}{\mcitedefaultseppunct}\relax
\EndOfBibitem
\bibitem[Sedzinski \emph{et~al.}(2011)Sedzinski, Biro, Oswald, Tinevez,
  Salbreux, and Paluch]{sedzinski2011}
J.~Sedzinski, M.~Biro, A.~Oswald, J.-Y. Tinevez, G.~Salbreux and E.~Paluch,
  \emph{Nature}, 2011, \textbf{476}, 462\relax
\mciteBstWouldAddEndPuncttrue
\mciteSetBstMidEndSepPunct{\mcitedefaultmidpunct}
{\mcitedefaultendpunct}{\mcitedefaultseppunct}\relax
\EndOfBibitem
\bibitem[Munro \emph{et~al.}(2004)Munro, Nance, and Priess]{munro2004}
E.~Munro, J.~Nance and J.~R. Priess, \emph{Dev. Cell}, 2004, \textbf{7},
  413--424\relax
\mciteBstWouldAddEndPuncttrue
\mciteSetBstMidEndSepPunct{\mcitedefaultmidpunct}
{\mcitedefaultendpunct}{\mcitedefaultseppunct}\relax
\EndOfBibitem
\bibitem[Tabei \emph{et~al.}(2013)Tabei, Burov, Kim, Kuznetsov, Huynh,
  Jureller, Philipson, Dinner, and Scherer]{tabei2013}
S.~M.~A. Tabei, S.~Burov, H.~Y. Kim, A.~Kuznetsov, T.~Huynh, J.~Jureller, L.~H.
  Philipson, A.~R. Dinner and N.~F. Scherer, \emph{Proc. Natl. Acad. Sci.
  U.S.A}, 2013, \textbf{110}, 4911--4916\relax
\mciteBstWouldAddEndPuncttrue
\mciteSetBstMidEndSepPunct{\mcitedefaultmidpunct}
{\mcitedefaultendpunct}{\mcitedefaultseppunct}\relax
\EndOfBibitem
\bibitem[Bendix \emph{et~al.}(2008)Bendix, Koenderink, Cuvelier, Dogic,
  Koeleman, Brieher, Field, Mahadevan, and Weitz]{bendix2008}
P.~M. Bendix, G.~H. Koenderink, D.~Cuvelier, Z.~Dogic, B.~N. Koeleman, W.~M.
  Brieher, C.~M. Field, L.~Mahadevan and D.~A. Weitz, \emph{Biophys. J.}, 2008,
  \textbf{94}, 3126--3136\relax
\mciteBstWouldAddEndPuncttrue
\mciteSetBstMidEndSepPunct{\mcitedefaultmidpunct}
{\mcitedefaultendpunct}{\mcitedefaultseppunct}\relax
\EndOfBibitem
\bibitem[Kohler \emph{et~al.}(2011)Kohler, Schaller,
  Bausch,\emph{et~al.}]{kohler2011}
S.~Kohler, V.~Schaller, A.~Bausch \emph{et~al.}, \emph{Nat. Mater.}, 2011,
  \textbf{10}, 462--468\relax
\mciteBstWouldAddEndPuncttrue
\mciteSetBstMidEndSepPunct{\mcitedefaultmidpunct}
{\mcitedefaultendpunct}{\mcitedefaultseppunct}\relax
\EndOfBibitem
\bibitem[Alvarado \emph{et~al.}(2013)Alvarado, Sheinman, Sharma, MacKintosh,
  and Koenderink]{alvarado2013}
J.~Alvarado, M.~Sheinman, A.~Sharma, F.~C. MacKintosh and G.~H. Koenderink,
  \emph{Nat Phys}, 2013, \textbf{9}, 591--597\relax
\mciteBstWouldAddEndPuncttrue
\mciteSetBstMidEndSepPunct{\mcitedefaultmidpunct}
{\mcitedefaultendpunct}{\mcitedefaultseppunct}\relax
\EndOfBibitem
\bibitem[Murrell and Gardel(2012)]{murrell2012}
M.~P. Murrell and M.~L. Gardel, \emph{Proc. Natl. Acad. Sci. USA}, 2012,
  \textbf{109}, 20820--20825\relax
\mciteBstWouldAddEndPuncttrue
\mciteSetBstMidEndSepPunct{\mcitedefaultmidpunct}
{\mcitedefaultendpunct}{\mcitedefaultseppunct}\relax
\EndOfBibitem
\bibitem[Murrell and Gardel(2014)]{murrell2014}
M.~Murrell and M.~L. Gardel, \emph{Mol. Biol. Cell}, 2014, \textbf{25},
  1845--1853\relax
\mciteBstWouldAddEndPuncttrue
\mciteSetBstMidEndSepPunct{\mcitedefaultmidpunct}
{\mcitedefaultendpunct}{\mcitedefaultseppunct}\relax
\EndOfBibitem
\bibitem[Ennomani \emph{et~al.}(2016)Ennomani, Letort, Gu{\'e}rin, Martiel,
  Cao, N{\'e}d{\'e}lec, Enrique, Th{\'e}ry, and Blanchoin]{ennomani2016}
H.~Ennomani, G.~Letort, C.~Gu{\'e}rin, J.-L. Martiel, W.~Cao,
  F.~N{\'e}d{\'e}lec, M.~Enrique, M.~Th{\'e}ry and L.~Blanchoin, \emph{Curr.
  Biol.}, 2016, \textbf{26}, 616--626\relax
\mciteBstWouldAddEndPuncttrue
\mciteSetBstMidEndSepPunct{\mcitedefaultmidpunct}
{\mcitedefaultendpunct}{\mcitedefaultseppunct}\relax
\EndOfBibitem
\bibitem[Linsmeier \emph{et~al.}(2016)Linsmeier, Banerjee, Oakes, Jung, Kim,
  and Murrell]{linsmeier2016}
I.~Linsmeier, S.~Banerjee, P.~W. Oakes, W.~Jung, T.~Kim and M.~Murrell,
  \emph{Nat. Commun.}, 2016, \textbf{7}, 12615\relax
\mciteBstWouldAddEndPuncttrue
\mciteSetBstMidEndSepPunct{\mcitedefaultmidpunct}
{\mcitedefaultendpunct}{\mcitedefaultseppunct}\relax
\EndOfBibitem
\bibitem[Chugh \emph{et~al.}(2017)Chugh, Clark, Smith, Cassani, Dierkes, Ragab,
  Roux, Charras, Salbreux, and Paluch]{chugh2017}
P.~Chugh, A.~G. Clark, M.~B. Smith, D.~A. Cassani, K.~Dierkes, A.~Ragab, P.~P.
  Roux, G.~Charras, G.~Salbreux and E.~K. Paluch, \emph{Nat. Cell Biol.}, 2017,
  \textbf{19}, 689--697\relax
\mciteBstWouldAddEndPuncttrue
\mciteSetBstMidEndSepPunct{\mcitedefaultmidpunct}
{\mcitedefaultendpunct}{\mcitedefaultseppunct}\relax
\EndOfBibitem
\bibitem[Stam \emph{et~al.}(2017)Stam, Freedman, Banerjee, Weirich, Dinner, and
  Gardel]{stam2017}
S.~Stam, S.~L. Freedman, S.~Banerjee, K.~L. Weirich, A.~R. Dinner and M.~L.
  Gardel, \emph{Proc. Natl. Acad. Sci. U.S.A}, 2017, \textbf{114},
  E10037--10045\relax
\mciteBstWouldAddEndPuncttrue
\mciteSetBstMidEndSepPunct{\mcitedefaultmidpunct}
{\mcitedefaultendpunct}{\mcitedefaultseppunct}\relax
\EndOfBibitem
\bibitem[Gardel \emph{et~al.}(2004)Gardel, Shin, MacKintosh, Mahadevan,
  Matsudaira, and Weitz]{gardel2004}
M.~Gardel, J.~Shin, F.~MacKintosh, L.~Mahadevan, P.~Matsudaira and D.~Weitz,
  \emph{Science}, 2004, \textbf{304}, 1301--1305\relax
\mciteBstWouldAddEndPuncttrue
\mciteSetBstMidEndSepPunct{\mcitedefaultmidpunct}
{\mcitedefaultendpunct}{\mcitedefaultseppunct}\relax
\EndOfBibitem
\bibitem[Kasza \emph{et~al.}(2009)Kasza, Koenderink, Lin, Broedersz, Messner,
  Nakamura, Stossel, MacKintosh, and Weitz]{kasza2009}
K.~E. Kasza, G.~H. Koenderink, Y.~C. Lin, C.~P. Broedersz, W.~Messner,
  F.~Nakamura, T.~P. Stossel, F.~C. MacKintosh and D.~A. Weitz, \emph{Phys.
  Rev. E}, 2009, \textbf{79}, 041928\relax
\mciteBstWouldAddEndPuncttrue
\mciteSetBstMidEndSepPunct{\mcitedefaultmidpunct}
{\mcitedefaultendpunct}{\mcitedefaultseppunct}\relax
\EndOfBibitem
\bibitem[Freedman \emph{et~al.}(2017)Freedman, Banerjee, Hocky, and
  Dinner]{freedman2017}
S.~L. Freedman, S.~Banerjee, G.~M. Hocky and A.~R. Dinner, \emph{Biophys. J},
  2017, \textbf{113}, 448--460\relax
\mciteBstWouldAddEndPuncttrue
\mciteSetBstMidEndSepPunct{\mcitedefaultmidpunct}
{\mcitedefaultendpunct}{\mcitedefaultseppunct}\relax
\EndOfBibitem
\bibitem[Schmoller \emph{et~al.}(2009)Schmoller, Lieleg, and
  Bausch]{schmoller2009}
K.~Schmoller, O.~Lieleg and A.~Bausch, \emph{Biophys. J.}, 2009, \textbf{97},
  83--89\relax
\mciteBstWouldAddEndPuncttrue
\mciteSetBstMidEndSepPunct{\mcitedefaultmidpunct}
{\mcitedefaultendpunct}{\mcitedefaultseppunct}\relax
\EndOfBibitem
\bibitem[Weirich \emph{et~al.}(2017)Weirich, Banerjee, Dasbiswas, Witten,
  Vaikuntanathan, and Gardel]{weirich2017}
K.~L. Weirich, S.~Banerjee, K.~Dasbiswas, T.~A. Witten, S.~Vaikuntanathan and
  M.~L. Gardel, \emph{Proc. Natl. Acad. Sci. U.S.A}, 2017, \textbf{114},
  2131--2136\relax
\mciteBstWouldAddEndPuncttrue
\mciteSetBstMidEndSepPunct{\mcitedefaultmidpunct}
{\mcitedefaultendpunct}{\mcitedefaultseppunct}\relax
\EndOfBibitem
\bibitem[Lieleg \emph{et~al.}(2009)Lieleg, Baumg{\"a}rtel, and
  Bausch]{lieleg2009}
O.~Lieleg, R.~M. Baumg{\"a}rtel and A.~R. Bausch, \emph{Biophys. J.}, 2009,
  \textbf{97}, 1569--1577\relax
\mciteBstWouldAddEndPuncttrue
\mciteSetBstMidEndSepPunct{\mcitedefaultmidpunct}
{\mcitedefaultendpunct}{\mcitedefaultseppunct}\relax
\EndOfBibitem
\bibitem[Scholz \emph{et~al.}(2018)Scholz, Weirich, Gardel, and
  Dinner]{scholz2018}
M.~Scholz, K.~L. Weirich, M.~L. Gardel and A.~R. Dinner, \emph{bioRxiv}, 2018,
  277947\relax
\mciteBstWouldAddEndPuncttrue
\mciteSetBstMidEndSepPunct{\mcitedefaultmidpunct}
{\mcitedefaultendpunct}{\mcitedefaultseppunct}\relax
\EndOfBibitem
\bibitem[Dasanayake \emph{et~al.}(2011)Dasanayake, Michalski, and
  Carlsson]{dasanyake2011}
N.~L. Dasanayake, P.~J. Michalski and A.~E. Carlsson, \emph{Phys. Rev. Lett.},
  2011, \textbf{107}, 118101\relax
\mciteBstWouldAddEndPuncttrue
\mciteSetBstMidEndSepPunct{\mcitedefaultmidpunct}
{\mcitedefaultendpunct}{\mcitedefaultseppunct}\relax
\EndOfBibitem
\bibitem[Belmonte \emph{et~al.}(2017)Belmonte, Leptin, and
  N{\'e}d{\'e}lec]{belmonte2017}
J.~M. Belmonte, M.~Leptin and F.~N{\'e}d{\'e}lec, \emph{Molecular Syst. Biol.},
  2017, \textbf{13}, 941\relax
\mciteBstWouldAddEndPuncttrue
\mciteSetBstMidEndSepPunct{\mcitedefaultmidpunct}
{\mcitedefaultendpunct}{\mcitedefaultseppunct}\relax
\EndOfBibitem
\bibitem[Head \emph{et~al.}(2003)Head, Levine, and MacKintosh]{head2003}
D.~A. Head, A.~J. Levine and F.~C. MacKintosh, \emph{Phys. Rev. E}, 2003,
  \textbf{68}, 061907\relax
\mciteBstWouldAddEndPuncttrue
\mciteSetBstMidEndSepPunct{\mcitedefaultmidpunct}
{\mcitedefaultendpunct}{\mcitedefaultseppunct}\relax
\EndOfBibitem
\bibitem[Kim \emph{et~al.}(2009)Kim, Hwang, Lee, and Kamm]{kim2009}
T.~Kim, W.~Hwang, H.~Lee and R.~D. Kamm, \emph{PLoS Comput. Biol.}, 2009,
  \textbf{5}, e1000439\relax
\mciteBstWouldAddEndPuncttrue
\mciteSetBstMidEndSepPunct{\mcitedefaultmidpunct}
{\mcitedefaultendpunct}{\mcitedefaultseppunct}\relax
\EndOfBibitem
\bibitem[Head \emph{et~al.}(2011)Head, Gompper, and Briels]{head2011}
D.~A. Head, G.~Gompper and W.~J. Briels, \emph{Soft Matter}, 2011, \textbf{7},
  3116--3126\relax
\mciteBstWouldAddEndPuncttrue
\mciteSetBstMidEndSepPunct{\mcitedefaultmidpunct}
{\mcitedefaultendpunct}{\mcitedefaultseppunct}\relax
\EndOfBibitem
\bibitem[Scholz \emph{et~al.}(2016)Scholz, Burov, Weirich, Scholz, Tabei,
  Gardel, and Dinner]{scholz2016}
M.~Scholz, S.~Burov, K.~L. Weirich, B.~J. Scholz, S.~M.~A. Tabei, M.~L. Gardel
  and A.~R. Dinner, \emph{Phys. Rev. X}, 2016, \textbf{6}, 011037\relax
\mciteBstWouldAddEndPuncttrue
\mciteSetBstMidEndSepPunct{\mcitedefaultmidpunct}
{\mcitedefaultendpunct}{\mcitedefaultseppunct}\relax
\EndOfBibitem
\bibitem[Cyron \emph{et~al.}(2013)Cyron, M{\"u}ller, Schmoller, Bausch, Wall,
  and Bruinsma]{cyron2013}
C.~Cyron, K.~M{\"u}ller, K.~Schmoller, A.~Bausch, W.~Wall and R.~Bruinsma,
  \emph{Europhys. Lett.}, 2013, \textbf{102}, 38003\relax
\mciteBstWouldAddEndPuncttrue
\mciteSetBstMidEndSepPunct{\mcitedefaultmidpunct}
{\mcitedefaultendpunct}{\mcitedefaultseppunct}\relax
\EndOfBibitem
\bibitem[Nedelec and Foethke(2007)]{nedelec2007}
F.~Nedelec and D.~Foethke, \emph{New J. Phys.}, 2007, \textbf{9}, 427\relax
\mciteBstWouldAddEndPuncttrue
\mciteSetBstMidEndSepPunct{\mcitedefaultmidpunct}
{\mcitedefaultendpunct}{\mcitedefaultseppunct}\relax
\EndOfBibitem
\bibitem[Popov \emph{et~al.}(2016)Popov, Komianos, and Papoian]{popov2016}
K.~Popov, J.~Komianos and G.~A. Papoian, \emph{PLoS Comput. Biol.}, 2016,
  \textbf{12}, e1004877\relax
\mciteBstWouldAddEndPuncttrue
\mciteSetBstMidEndSepPunct{\mcitedefaultmidpunct}
{\mcitedefaultendpunct}{\mcitedefaultseppunct}\relax
\EndOfBibitem
\bibitem[Mickel \emph{et~al.}(2008)Mickel, M{\"u}nster, Jawerth, Vader, Weitz,
  Sheppard, Mecke, Fabry, and Schr{\"o}der-Turk]{mickel2008}
W.~Mickel, S.~M{\"u}nster, L.~M. Jawerth, D.~A. Vader, D.~A. Weitz, A.~P.
  Sheppard, K.~Mecke, B.~Fabry and G.~E. Schr{\"o}der-Turk, \emph{Biophys. J.},
  2008, \textbf{95}, 6072--6080\relax
\mciteBstWouldAddEndPuncttrue
\mciteSetBstMidEndSepPunct{\mcitedefaultmidpunct}
{\mcitedefaultendpunct}{\mcitedefaultseppunct}\relax
\EndOfBibitem
\bibitem[Fritzsche \emph{et~al.}(2016)Fritzsche, Erlenk{\"a}mper, Moeendarbary,
  Charras, and Kruse]{fritzsche2016}
M.~Fritzsche, C.~Erlenk{\"a}mper, E.~Moeendarbary, G.~Charras and K.~Kruse,
  \emph{Sci. Adv.}, 2016, \textbf{2}, e1501337\relax
\mciteBstWouldAddEndPuncttrue
\mciteSetBstMidEndSepPunct{\mcitedefaultmidpunct}
{\mcitedefaultendpunct}{\mcitedefaultseppunct}\relax
\EndOfBibitem
\bibitem[Mohapatra \emph{et~al.}(2016)Mohapatra, Goode, Jelenkovic, Phillips,
  and Kondev]{mohapatra2016}
L.~Mohapatra, B.~L. Goode, P.~Jelenkovic, R.~Phillips and J.~Kondev,
  \emph{Annu. Rev. Biophys.}, 2016, \textbf{45}, 85--116\relax
\mciteBstWouldAddEndPuncttrue
\mciteSetBstMidEndSepPunct{\mcitedefaultmidpunct}
{\mcitedefaultendpunct}{\mcitedefaultseppunct}\relax
\EndOfBibitem
\bibitem[Chesarone-Cataldo \emph{et~al.}(2011)Chesarone-Cataldo, Gu{\'e}rin,
  Jerry, Wedlich-Soldner, Blanchoin, and Goode]{chesarone2011}
M.~Chesarone-Cataldo, C.~Gu{\'e}rin, H.~Y. Jerry, R.~Wedlich-Soldner,
  L.~Blanchoin and B.~L. Goode, \emph{Dev. Cell}, 2011, \textbf{21},
  217--230\relax
\mciteBstWouldAddEndPuncttrue
\mciteSetBstMidEndSepPunct{\mcitedefaultmidpunct}
{\mcitedefaultendpunct}{\mcitedefaultseppunct}\relax
\EndOfBibitem
\bibitem[Lin \emph{et~al.}(2005)Lin, Schneider, and Kachar]{lin2005}
H.~W. Lin, M.~E. Schneider and B.~Kachar, \emph{Curr. Opin. Cell Biol.}, 2005,
  \textbf{17}, 55--61\relax
\mciteBstWouldAddEndPuncttrue
\mciteSetBstMidEndSepPunct{\mcitedefaultmidpunct}
{\mcitedefaultendpunct}{\mcitedefaultseppunct}\relax
\EndOfBibitem
\bibitem[Salbreux \emph{et~al.}(2012)Salbreux, Charras, and
  Paluch]{salbreux2012}
G.~Salbreux, G.~Charras and E.~Paluch, \emph{Trends Cell Biol.}, 2012,
  \textbf{22}, 536--545\relax
\mciteBstWouldAddEndPuncttrue
\mciteSetBstMidEndSepPunct{\mcitedefaultmidpunct}
{\mcitedefaultendpunct}{\mcitedefaultseppunct}\relax
\EndOfBibitem
\bibitem[Brawley and Rock(2009)]{brawley2009}
C.~M. Brawley and R.~S. Rock, \emph{Proc. Natl. Acad. Sci. U.S.A}, 2009,
  \textbf{106}, 9685--9690\relax
\mciteBstWouldAddEndPuncttrue
\mciteSetBstMidEndSepPunct{\mcitedefaultmidpunct}
{\mcitedefaultendpunct}{\mcitedefaultseppunct}\relax
\EndOfBibitem
\bibitem[Burov \emph{et~al.}(2013)Burov, Tabei, Huynh, Murrell, Philipson,
  Rice, Gardel, Scherer, and Dinner]{burov2013}
S.~Burov, S.~M.~A. Tabei, T.~Huynh, M.~P. Murrell, L.~H. Philipson, S.~A. Rice,
  M.~L. Gardel, N.~F. Scherer and A.~R. Dinner, \emph{Proc. Natl. Acad. Sci.
  USA}, 2013, \textbf{110}, 19689--19694\relax
\mciteBstWouldAddEndPuncttrue
\mciteSetBstMidEndSepPunct{\mcitedefaultmidpunct}
{\mcitedefaultendpunct}{\mcitedefaultseppunct}\relax
\EndOfBibitem
\bibitem[Sanchez \emph{et~al.}(2012)Sanchez, Chen, DeCamp, Heymann, and
  Dogic]{sanchez2012}
T.~Sanchez, D.~T. Chen, S.~J. DeCamp, M.~Heymann and Z.~Dogic, \emph{Nature},
  2012, \textbf{491}, 431\relax
\mciteBstWouldAddEndPuncttrue
\mciteSetBstMidEndSepPunct{\mcitedefaultmidpunct}
{\mcitedefaultendpunct}{\mcitedefaultseppunct}\relax
\EndOfBibitem
\bibitem[Foster \emph{et~al.}(2015)Foster, F{\"u}rthauer, Shelley, and
  Needleman]{foster2015}
P.~J. Foster, S.~F{\"u}rthauer, M.~J. Shelley and D.~J. Needleman,
  \emph{eLife}, 2015, \textbf{4}, e10837\relax
\mciteBstWouldAddEndPuncttrue
\mciteSetBstMidEndSepPunct{\mcitedefaultmidpunct}
{\mcitedefaultendpunct}{\mcitedefaultseppunct}\relax
\EndOfBibitem
\end{mcitethebibliography}
\providecommand*{\mcitethebibliography}{\thebibliography}
\csname @ifundefined\endcsname{endmcitethebibliography}
{\let\endmcitethebibliography\endthebibliography}{}

\setboolean{@twoside}{false}
\setlength{\voffset}{0cm}
\setlength{\hoffset}{0cm}


\section*{Supplementary material}

# Supplemental Materials for
# Design principles for selective self-assembly of active networks

Simon L. Freedman, Glen M. Hocky, Shiladitya Banerjee, Aaron R. Dinner

# S1 AFINES simulation

In AFINES, actin filaments, myosin motors, and passive crosslinker proteins are modeled as coarse grained entities. Actin filaments are treated as worm-like chains of $N+1$ beads connected by $N$ harmonic springs (links) and $N-1$ angular harmonic springs. Thus, the internal forces on an actin filament can be obtained from the gradient of the potential energy $U_f$:

$$U_f = \frac{k_a}{2} \sum_{i=1}^{N} (|\vec{r}_i - \vec{r}_{i-1}| - l_a)^2 + \frac{\kappa_B}{2l_a} \sum_{i=2}^{N} \theta_i^2, \tag{1}$$

where $\vec{r}_i$ is the position of the $i^{th}$ bead on a filament, $\theta_i$ is the angle between the $i^{th}$ and $(i-1)^{th}$ links, $k_a$ is the stretching force constant, $\kappa_B$ is the bending modulus, and $l_a$ is the equilibrium length of a link. The persistence length of the filament is then $L_p = \kappa_B/k_BT$, where $k_B$ is Boltzmann's constant and $T$ is the temperature.

We model crosslinkers as Hookean springs, with two ends (heads) that can stochastically bind to and unbind from filaments. Thus, the potential energy of a crosslinker is

$$U_{xl} = \frac{1}{2} k_{xl} (|\vec{r}_1 - \vec{r}_2| - l_{xl})^2 - k_BT \ln (k_{xl}^{\text{on}}/k_{xl}^{\text{off}})(I_1 + I_2) \tag{2}$$

where $k_{xl}$ is the crosslinker stiffness, $l_{xl}$ is its rest length, $\vec{r}_{1(2)}$ is the position of head 1(2), $I_{1(2)}$ is 1 if head 1(2) is bound and 0 otherwise, and $k_{xl}^{\text{on}}(k_{xl}^{\text{off}})$ is the rate constant for binding (unbinding). When a crosslinker is bound, it moves with the filament to which it is bound. When both crosslinker heads are bound, its tensile force, $\vec{F}_{xl}$ is propagated onto the filament beads neighboring each bound head at position $\vec{r}_{xl}$ via the lever rule,

$$\vec{F}_i = \vec{F}_{xl} \frac{|\vec{r}_{i+1} - \vec{r}_{xl}|}{|\vec{r}_{i+1} - \vec{r}_i|} \quad \text{and} \quad \vec{F}_{i+1} = \vec{F}_{xl} - \vec{F}_i, \tag{3}$$

where $\vec{F}_i$ is the force on the filament bead at position $\vec{r}_i$.

Binding and unbinding are governed by a Monte Carlo procedure constructed to satisfy detailed balance in the absence of motors. At each timestep of duration $\Delta t$, an unbound crosslinker head becomes bound to the $i^{th}$ nearby filament with probability $k_{xl}^{\text{on}} \Delta t P_i$, where $P_i$ is defined as follows. The closest point on the filament is identified, and the change in energy

<div align="center">1</div>

associated with moving the head to it, $\Delta U_i$ is computed; $P_i = \min\left[1, \exp\left(-\Delta U_i/k_B T\right)\right]$. When a head becomes bound, its displacement, in the frame of reference of the filament link to which it attached, is stored as $\Delta \vec{r}$. Later, the head can become unbound with probability $k_{xl}^{\text{off}} \Delta t P$, where $P = \min\left[1, \exp(-\Delta U/k_B T)\right]$ and $\Delta U$ is the energy that would result from applying a displacement of $-\Delta \vec{r}$ to the head.

We model a motor similarly to a crosslinker, in that it is a Hookean spring, can bind to and unbind from filaments, and propagates force onto them. Thus, their potential energy is identical to Eq. 2 with the subscript $m$ replacing the subscript $xl$. Additionally, a bound motor head moves towards the barbed end of the actin filament to which it is bound at a load dependent velocity

$$v(F_m) = v_0 \max\left[1 + \frac{\vec{F}_m \cdot \hat{r}}{F_s}, 0\right],\tag{4}$$

where $v_0$ is the unloaded motor speed, $\vec{F}_m = -k_m(|\vec{r}_1 - \vec{r}_2| - l_m)$ is the tensile force on the motor, and $\hat{r}$ is the tangent to the filament at the point where the motor is bound; $\hat{r}$ points toward the pointed end of the filament.

We simulate the system using Brownian dynamics such that the position of an actin bead, motor head, or crosslinker head at time $t$ is generated by the equation

$$\vec{r}(t + \Delta t) = \vec{r}(t) + \vec{F}(\vec{r}(t))\mu \Delta t + \sqrt{\frac{k_B T \mu \Delta t}{2}}\left(\vec{W}(t + \Delta t) + \vec{W}(t)\right),\tag{5}$$

where $\vec{F}(\vec{r}(t))$ is the gradient of the potential of the particle, $\vec{W}(t)$ is a vector of random numbers drawn from the standard normal distribution, and we use the Stokes relation $\mu = 1/(6\pi R \nu)$ in the damping term, where $R$ is the size of the particle, and $\nu$ is the dynamic viscosity of its environment [S1]. We simulate the system in 2D and use periodic boundary conditions to limit boundary effects. A complete list of model parameters used for Figures 1, 2, 4 and 5 is provided in Table S1 and additional methods used for Fig. 8 are described below (Section S3).

| Symbol | Description (units) [ref] | Value |
|--------|---------------------------|-------|
| | **Actin Filaments** | |
| $\rho_l$ | link density ($\mu m^{-2}$) | 2 |
| $N_l$ | number of links per filament ($L/l_a$) | $[1, 15]$ |
| $l_a$ | link rest length ($\mu m$) [S2] | 1 |
| $k_a$ | stretching force constant (pN/$\mu m$) | 5 |
| $\kappa_B$ | bending modulus (pN$\mu m^2$) [S3] | 0.068 |
| | **Myosin Minifilaments** | |
| $\rho_m$ | density ($\mu m^{-2}$) | $[0, 0.3]$ |
| $l_m$ | rest length ($\mu m$) [S4] | 0.5 |
| $k_m$ | stiffness (pN/$\mu m$) | 1 |
| $k_m^{\mathrm{on}}$ | maximum attachment rate (s$^{-1}$) | 1 |
| $k_m^{\mathrm{off}}$ | maximum detachment rate (s$^{-1}$) | $[0.01, 1]$ |
| $k_m^{\mathrm{end}}$ | maximum end detachment rate (s$^{-1}$) | $[0.01, 1]$ |
| $v_0$ | unloaded speed ($\mu m/s$) [S5] | 1 |
| $F_s$ | stall force of myosin (pN) [S6] | 0.5 |
| | **Crosslinkers** | |
| $\rho_{xl}$ | density ($\mu m^{-2}$) | $[0, 1.5]$ |
| $l_{xl}$ | rest length (filamin) ($\mu m$) [S7] | 0.15 |
| $k_{xl}$ | stiffness (pN/$\mu m$) | 1 |
| $k_{xl}^{\mathrm{on}}$ | maximum attachment rate (s$^{-1}$) | 1 |
| $k_{xl}^{\mathrm{off}}$ | maximum detachment rate (s$^{-1}$) | $[0.01, 1]$ |
| | **Environment** | |
| $\Delta t$ | dynamics timestep (s) | 0.00002 |
| $t_F$ | maximum simulated time (s) | 400 |
| $X, Y$ | length and width of assay ($\mu m$) | 50 |
| $g$ | grid density ($\mu m^{-1}$) | 2.5 |
| $T$ | temperature ($K$) | 300 |
| $\nu$ | dynamic viscosity (Pa·s) | 0.001 |

Table S1: Parameter Values

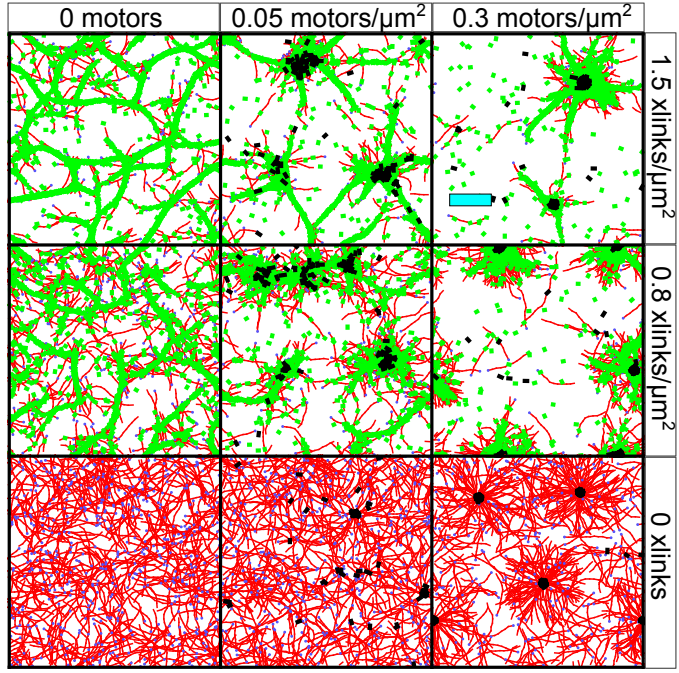

Figure S1: Same as Fig. 1, but with motors (black) and crosslinkers (green) also shown on top of actin (red, barbed end marked by blue dot). Cyan scale bar is 10 $\mu$m. Parameters held constant: $L = 10$ $\mu$m, $k_{xl(m)}^{\text{off}} = 0.1$ s$^{-1}$.

## S2 Actin velocity field calculation

Contraction of actin networks is typically calculated by identifying sinks in the divergence of the actin velocity field. To construct this field, we first calculate the velocity of actin beads $\vec{v}_a = (\vec{r}_a(t+h) - \vec{r}_a(t))/h$ where $\vec{r}_a(t)$ is the position of an actin bead at time $t$, and $h$ is the lag time. To reduce noise, we calculate $\vec{v}_k(\vec{r}_k)$, the average velocity in every bin $k$ of size $\Delta r^2$. We then interpolate the velocity field using Gaussian Radial Basis Functions (RBF), such that the velocity at any position $\vec{r}$ is

$$\vec{v}(\vec{r}) = \sum_{k=1}^{M} \vec{w}_k e^{-(|\vec{r}-\vec{r}_k|/\epsilon)^2}, \tag{6}$$

where $M$ is the number of bins with at least 10 actin beads, and $\vec{w}_k$ are the weights of the basis functions, determined by solving the equation $\vec{v}(\vec{r}_k) = \vec{v}_k(\vec{r}_k)$ (using the scipy.interpolate.Rbf package [S8]). For Fig. 2B, we used a lag time of $h = 10$ s, a threshold of $n = 10$ actin beads in a local box of size $\Delta r = 5$ $\mu$m and a Gaussian width of $\epsilon = 5$ $\mu$m, as we have found these interpolation values robustly capture the motion of the actin [S9].

Because of the periodic boundary conditions, there is no flux of actin into the simulation cell, and $\langle \nabla \cdot \vec{v}(r) \rangle = 0$. Therefore, to measure contraction, we threshold the divergence by the local actin density and only total the divergence from 1 $\mu$m$^2$ patches that contain more than 5 actin beads in Fig. 2D, as in Ref. S10.

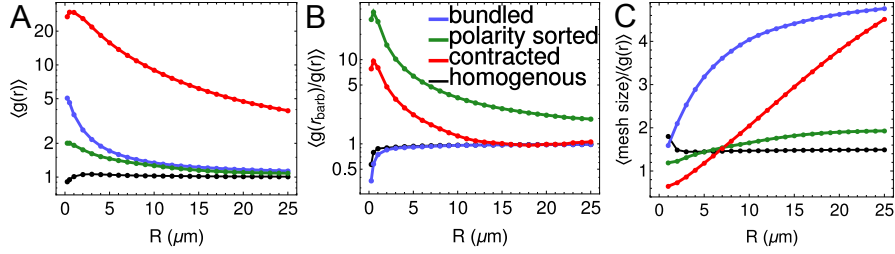

Figure S2: Effect of varying $R$, the upper integration limit for $\langle g(r) \rangle$ and $\langle g(r_{barb}) \rangle$, on order parameters. Since the length and width of the periodic simulation cell is 50 $\mu$m, the largest unambiguous distance is half that, 25 $\mu$m. (A) For all values of $R$, $\langle g(r) \rangle$ is highest for contracted networks, followed by bundled networks. (B) For all values of $R$, $\langle g(r_{barb}) / g(r) \rangle$ is highest for polarity sorted networks, followed by contracted networks. (C) For $R > 1$ $\mu$m, $\langle \text{meshsize}/g(r) \rangle$ is highest for bundled networks, consistently followed by contracted networks for $R \geq 8$ $\mu$m.

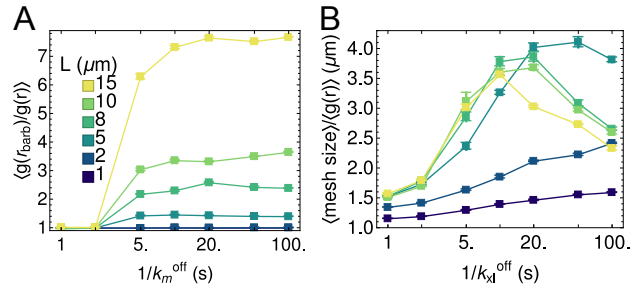

Figure S3: Same as Fig. 5B-C but normalized by $\langle g(r) \rangle$

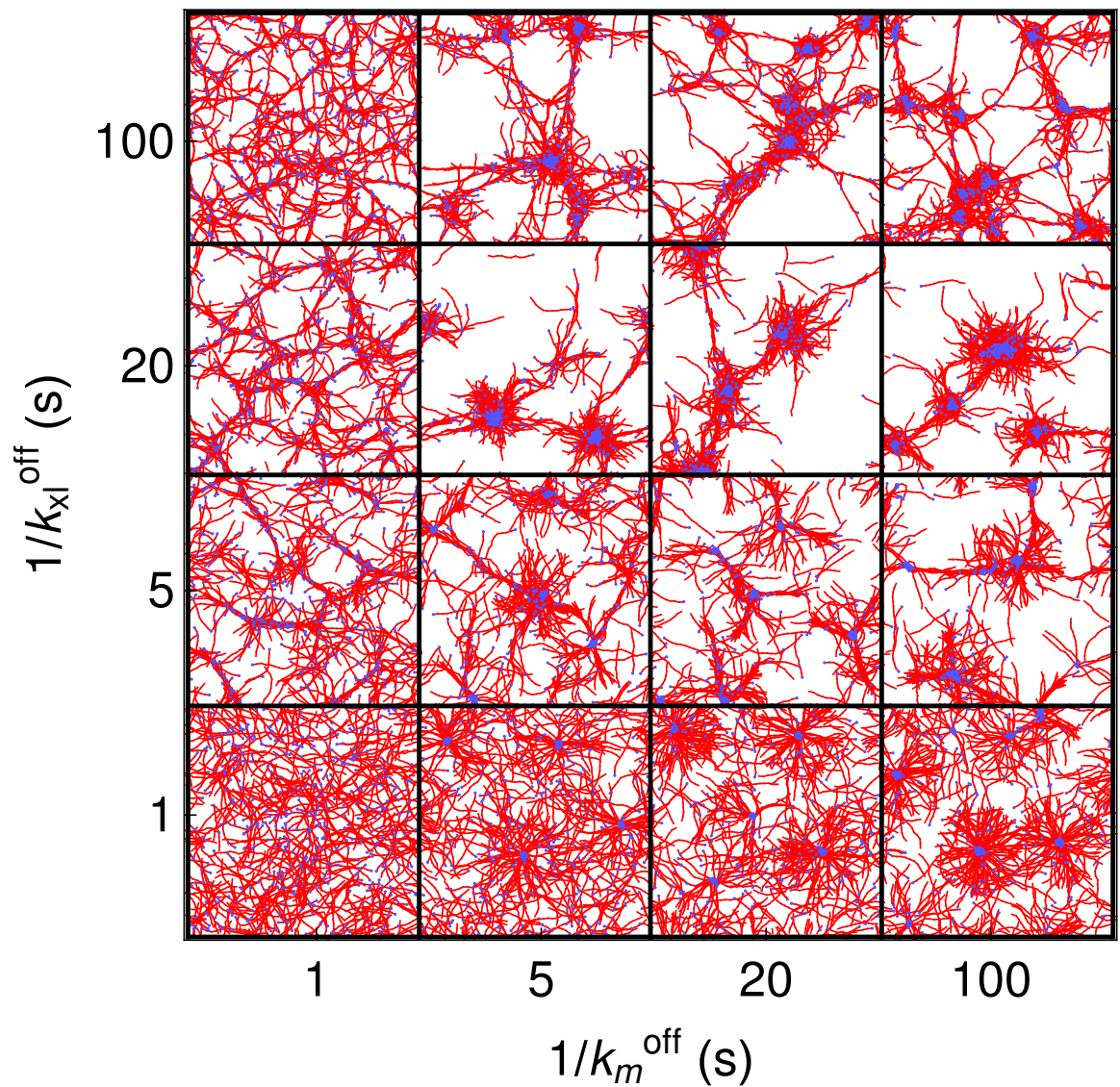

Figure S4: Sample structures after 400 s simulation, for varying motor and crosslinker off rates. Parameters held constant: $\rho_m = 0.2~\mu\text{m}^{-2}$, $\rho_{xl} = 1~\mu\text{m}^{-2}$, $L = 10~\mu\text{m}$. Motors and crosslinkers not shown for clarity.

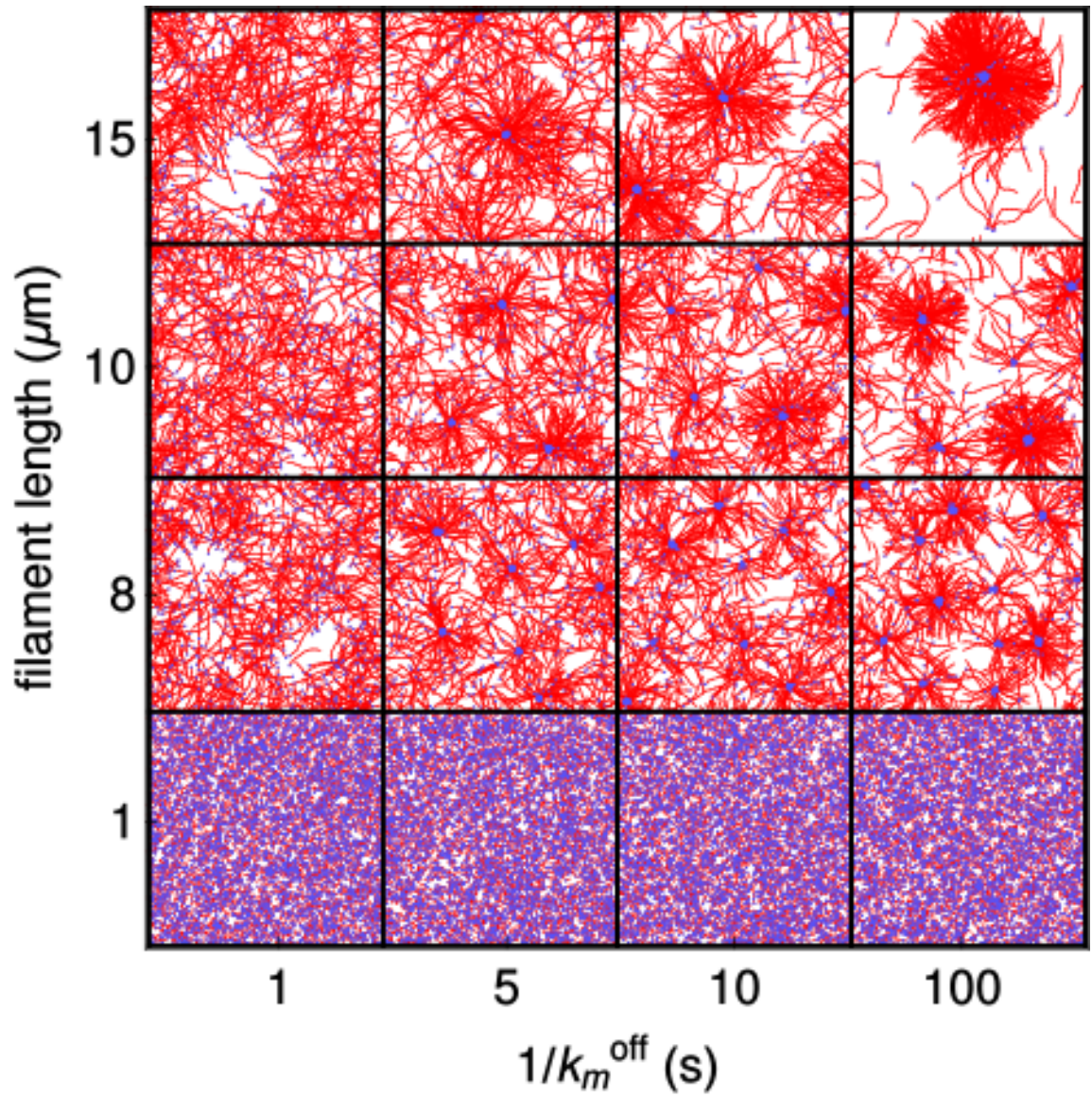

Figure S5: Sample structures after 400 s simulation, for varying filament length and motor off rate. Parameters held constant: $\rho_m = 0.3\ \mu\mathrm{m}^{-2}$, $\rho_{xl} = 0$. Motors and crosslinkers not shown for clarity.

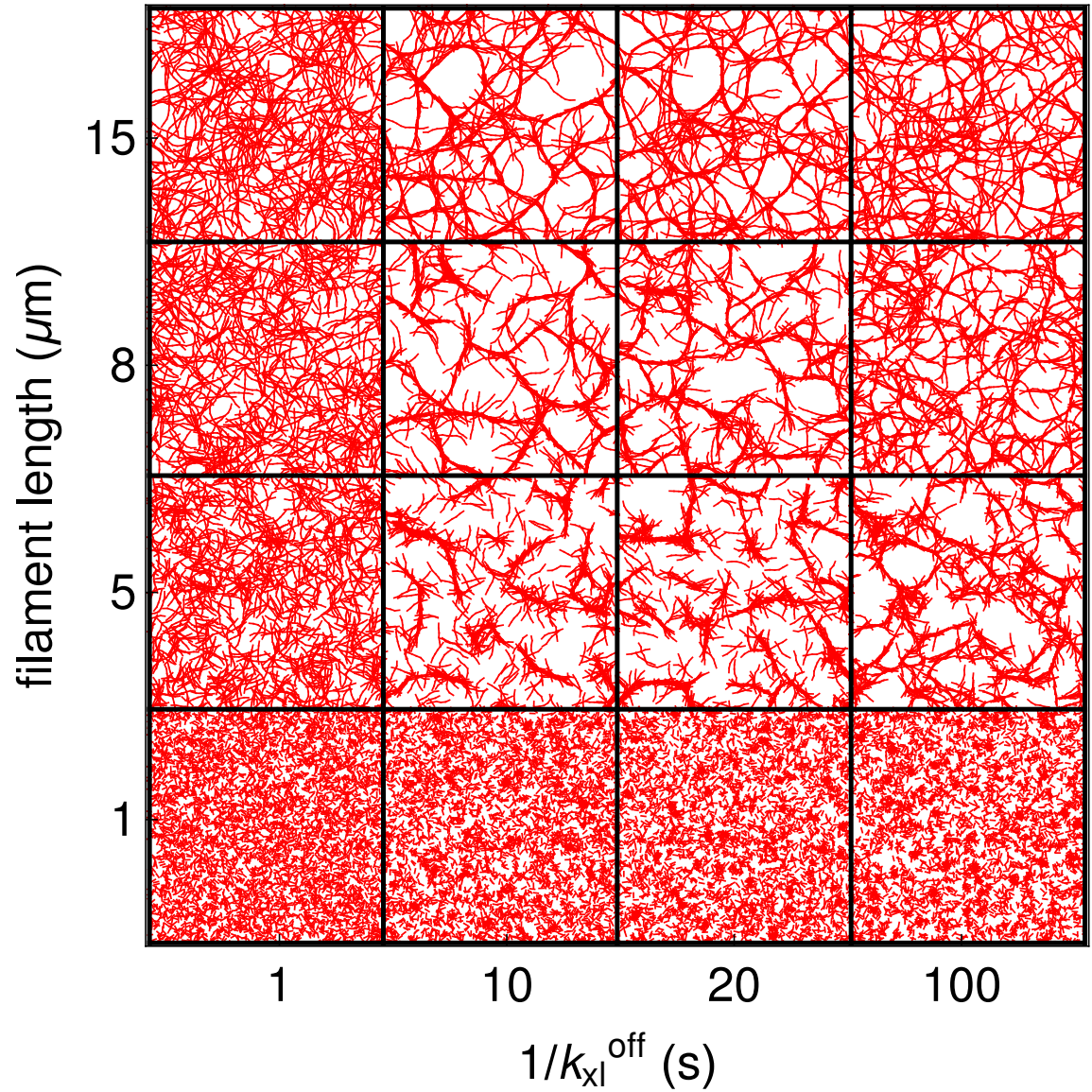

Figure S6: Sample structures after 400 s simulation, for varying filament length and crosslinker off rate. Parameters held constant: $\rho_m = 0$, $\rho_{xl} = 1.5\ \mu\mathrm{m}^{-2}$. Motors, crosslinkers, and barbed ends not shown for clarity.

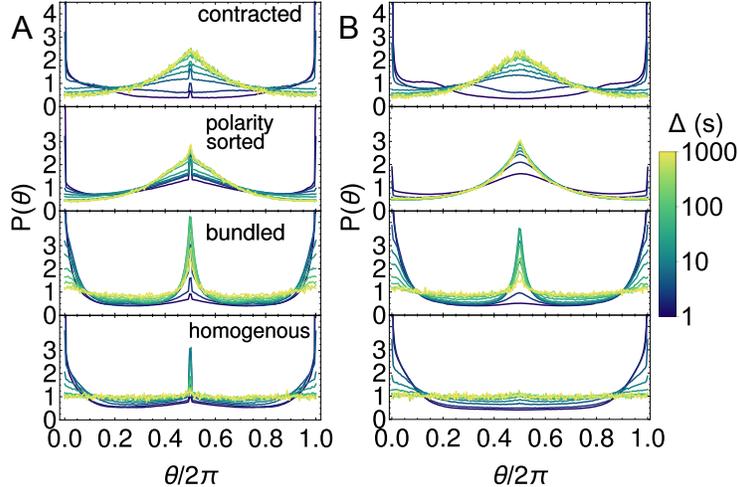

Figure S7: Caging of motors, measured using the distribution of angles between velocity vectors [S11]. (A) Caging at all time scales for all structures for motors with $k_m^{\text{end}} = k_m^{\text{off}}$. (B) Same as (A) but for motors with $k_m^{\text{end}} = 100 k_m^{\text{off}}$.

## S3 Simulating shear

To measure the stiffness of self-assembled actin networks, we simulate a controlled strain experiment and shear the final network configuration by a total strain of $\gamma = 0.5$ in a fixed amount of time $t_F = 0.5$ s. This is accomplished by supplementing the Brownian Dynamics described in Section S1 by explicitly shifting the actin bead position $(x_i, y_i)$ such that $x_i \rightarrow x_i + \gamma(dt/t_F)(y_i/Y)$ where $Y$ is the simulation cell height, and $dt$ is the amount of time for a small shear [S12]. Additionally, the boundary conditions follow the Lees-Edwards convention during the shear [S13]. As described in [S9], we do not perform this shift at every time step; rather $dt = \Delta t + t_{relax}$ where $\Delta t$ is the simulation time step and $t_{relax}$ is a suitable amount of time for the simulation to relax from the large external force imposed by the shear. In Fig. 2F, we used $\Delta t = 10^{-7}$ s and $dt = 10^{-3}$ s.

## References

[S1] Benedict Leimkuhler and Charles Matthews. Robust and efficient configurational molecular sampling via langevin dynamics. *J. Chem. Phys.*, 138(17):174102, 2013.

[S2] Theo Odijk. The statistics and dynamics of confined or entangled stiff polymers. *Macromolecules*, 16(8):1340–1344, 1983.

[S3] A. Ott, M. Magnasco, A. Simon, and A. Libchaber. Measurement of the persistence length of polymerized actin using fluorescence microscopy. *Phys. Rev. E*, 48:R1642–R1645, Sep 1993.

[S4]  Richard Niederman and Thomas D. Pollard. Human platelet myosin ii in vitro assembly and structure of myosin filaments. *J. Cell Biol.*, 67(1):72–92, 1975.

[S5]  Stephen J. Kron and James A. Spudich. Fluorescent actin filaments move on myosin fixed to a glass surface. *Proc. Natl. Acad. Sci. USA*, 83(17):6272–6276, 1986.

[S6]  Claudia Veigel, Justin E. Molloy, Stephan Schmitz, and John Kendrick-Jones. Load-dependent kinetics of force production by smooth muscle myosin measured with optical tweezers. *Nat. Cell Biol.*, 5(11):980–986, 2003.

[S7]  Jorge M. Ferrer, Hyungsuk Lee, Jiong Chen, Benjamin Pelz, Fumihiko Nakamura, Roger D. Kamm, and Matthew J. Lang. Measuring molecular rupture forces between single actin filaments and actin-binding proteins. *Proc. Natl. Acad. Sci. USA*, 105(27):9221–9226, 2008.

[S8]  Robert Hetland and John Travers. SciPy: Open source scientific tools for Python: rbf - radial basis functions for interpolation/smoothing scattered nd data, 2001.

[S9]  Simon L Freedman, Shiladitya Banerjee, Glen M Hocky, and Aaron R Dinner. A versatile framework for simulating the dynamic mechanical structure of cytoskeletal networks. *Biophys. J*, 113(2):448–460, 2017.

[S10] Samantha Stam, Simon L. Freedman, Shiladitya Banerjee, Kimberly L. Weirich, Aaron R. Dinner, and Margaret L. Gardel. Filament rigidity and connectivity tune the deformation modes of active biopolymer networks. *Proc. Natl. Acad. Sci. U.S.A*, 114(47):E10037–10045, 2017.

[S11] Stanislav Burov, S.M. Ali Tabei, Toan Huynh, Michael P. Murrell, Louis H. Philipson, Stuart A. Rice, Margaret L. Gardel, Norbert F. Scherer, and Aaron R. Dinner. Distribution of directional change as a signature of complex dynamics. *Proc. Natl. Acad. Sci. USA*, 110(49):19689–19694, 2013.

[S12] Denis J. Evans and G. P. Morriss. Nonlinear-response theory for steady planar couette flow. *Phys. Rev. A*, 30:1528–1530, Sep 1984.

[S13] AW Lees and SF Edwards. The computer study of transport processes under extreme conditions. *J. Phys. C: Solid State Phys.*, 5(15):1921, 1972.


\setlength{\voffset}{-2.54cm}
\setlength{\hoffset}{-2.54cm}
\end{document}